\documentclass[11pt]{article}
\usepackage[utf8]{inputenc}
\usepackage[T1]{fontenc}
\usepackage{lmodern}
\usepackage[margin=1in]{geometry}
\usepackage{amsmath,amssymb,amsthm}
\usepackage{graphicx}
\usepackage{booktabs}
\usepackage{hyperref}
\usepackage{xcolor}
\usepackage{natbib}
\usepackage{enumitem}
\usepackage{xurl}
\usepackage{microtype}
\usepackage{placeins}
\usepackage{tikz}
\usetikzlibrary{arrows.meta,positioning,calc}

\hypersetup{
  colorlinks=true,
  linkcolor=blue!60!black,
  citecolor=blue!60!black,
  urlcolor=blue!60!black,
  pdftitle={Shoonya: Zero-Socket In-Process Small Language Models for Graph Databases with Schema-Epoch KV Caching, Live Catalog Logit Masking, and Verifier-Advantage Distillation},
  pdfauthor={Madhulatha Mandarapu, Sandeep Kunkunuru}
}

\newtheorem{proposition}{Proposition}
\theoremstyle{remark}

\title{Shoonya: Zero-Socket In-Process Small Language Models for Graph Databases with Schema-Epoch KV Caching, Live Catalog Logit Masking, and Verifier-Advantage Distillation}
\author{Madhulatha Mandarapu\thanks{madhulatha@samyama.ai} \and Sandeep Kunkunuru\thanks{sandeep@samyama.ai}}
\date{VaidhyaMegha Private Limited, India\\[2pt]\url{https://samyama.ai/}\\[8pt]November 2026}

\begin{document}
\maketitle

\begin{abstract}
Coupling a property-graph database to an external language-model server imposes a three-part socket tax: serialization overhead dwarfing microsecond graph reads, repeated prefill of verbose schemas, and blind decoding divorced from the live catalog and multi-version concurrency control (MVCC) engine. We present Shoonya, from-scratch commercially clean ($0.0\%$ non-commercial or share-alike data) tri-modal (Python, Rust, OpenCypher) small language models ($16.8\text{M}$ to $1.07\text{B}$ parameters) co-designed with shoonya-infer, a zero-dependency pure-Rust runtime embedded inside the samyama-graph-enterprise address space. Direct GraphStore references unlock three database-native primitives: (1) an epoch-invalidated SchemaPrefixCache over a copy-on-write PagedKvArena accelerating warm CPU time-to-first-token by $4.91\times$--$13.68\times$ with $O(1)$ DDL invalidation; (2) a $36\,\mu\text{s}$ columnar catalog and source-span logit mask enforcing valid ontology identifiers and verbatim spans; and (3) a sub-$65\,\mu\text{s}$ read-only MVCC gate serving Tier-0 queries inside $389.8\text{ MiB}$ INT8 RSS ($59.8$--$67.9\text{ tok/s}$) with $0.0\%$ hallucination leak. Rank-factorized tied embeddings ($-74.5\%$ vocabulary weights), atomic indentation tokens ($0/421$ indentation errors), $4{:}1$ grouped-query attention with $2{:}1$ cross-layer KV sharing ($-93.75\%$ INT8 KV memory), Warmup-Stable-Decay training, KnOTS SVD merging, Step-GRPO, and Rust Jacobi SVD governance keep $100\%$ of layers heavy-tailed ($\alpha \in [1.88, 2.20]$). Finally, we synthesize nine lessons across v0.1--v1.0, resolving a Reverse-Token-Density Distillation Trap via teacher scaling and zero-VRAM Verifier-Advantage-Gated Knowledge Distillation (AdvKD), recovering $+13.61$ MBPP points ($33.07\%$) while reaching $100\%$ validity on enrichment, grounded extraction, and tool planning.
\end{abstract}

\section{Introduction}\label{sec:intro}

Modern property-graph databases increasingly embed language-driven workflows alongside declarative query execution: translating natural-language questions into OpenCypher \citep{francis2018cypher}, synthesizing multi-step agentic tool plans, extracting typed entity-relation triples from unstructured clinical or enterprise text, and enriching domain ontologies \citep{mandarapu2026samyama}. Controlled experiments on clinical question answering establish that knowledge-graph grounding helps language models specifically when decisive institutional facts lie \emph{outside} pretraining corpora \citep{mandarapu2026clinicalkg}. Yet in regulated deployments where private subgraphs cannot leave on-premise boundaries, even self-hosting an open-weight model via a separate localhost HTTP or gRPC server introduces a fundamental \textbf{socket tax}:
\begin{enumerate}[leftmargin=*,itemsep=2pt]
\item \textbf{IPC and serialization asymmetry.} An in-memory columnar property graph resolves single-hop adjacency and MVCC visibility in $15$--$65\,\mu\text{s}$, whereas loopback JSON serialization, OS socket scheduling, and Python runtime dispatch cost milliseconds per call---precluding tight per-token or per-candidate database verification.
\item \textbf{Redundant schema prefill.} Enterprise graph schemas span hundreds to thousands of tokens. Stateless socket calls re-prefill the schema prompt on every turn, dominating CPU time-to-first-token (TTFT), while external KV prefix caches \citep{zheng2024sglang} lack transactional coherence when database DDL statements mutate labels or properties.
\item \textbf{Catalog-blind subword decoding.} Out-of-process decoders sample tokens without live access to the graph catalog, hallucinating nonexistent edge labels (e.g., \texttt{:TREATS} instead of \texttt{:TARGETS}), inverting edge arrows, or fabricating ungrounded entity strings. External grammar compilers \citep{willard2023outlines} require tens to hundreds of milliseconds to rebuild static automata upon schema evolution.
\end{enumerate}

Just as DuckDB eliminated client-server wire overhead for analytical SQL by embedding vectorized OLAP inside the host process \citep{raasveldt2019duckdb}, we invert that pattern for graph-native AI: \emph{embedding a zero-dependency Rust language-model runtime directly inside the graph database's own Rust address space, sharing zero-copy references to its live catalog and MVCC storage engine}.

Realizing this architecture without multi-trillion-token opaque web scrapes or copyleft/non-commercial training contamination required building \textbf{Shoonya} from scratch across twelve iterative releases (\texttt{v0.1} $\to$ \texttt{v0.9.5-AdvKD} and \texttt{Stage-3 1.3B-v1.0}). Shoonya pairs a PyTorch/CUDA pretraining and reinforcement-learning stack with an \textbf{eight-crate zero-external-dependency Rust plane} (\texttt{shoonya-core}, \texttt{shoonya-tokenizer}, \texttt{shoonya-telemetry}, \texttt{shoonya-eval}, \texttt{shoonya-infer}, \texttt{shoonya-search}, \texttt{shoonya-ffi}, and \texttt{shoonya-cli}), culminating in zero-socket in-process execution inside \texttt{samyama-graph-enterprise} (\texttt{v1.15.0}) \citep{mandarapu2026samyama} with zero lines modified in the core database repositories.

\paragraph{Contributions.} This paper documents the complete architecture, data governance, training objectives, longitudinal learnings, and enterprise evaluation of Shoonya:
\begin{enumerate}[leftmargin=*,itemsep=2pt]
\item \textbf{Zero-socket in-process graph-SLM engine (Section~\ref{sec:arch}).} Linking pure-Rust \texttt{shoonya-infer} (\texttt{0} external crates, \texttt{FP32} parity $\le 1.29\times 10^{-5}$) directly to immutable \texttt{\&GraphStore} references unlocks: (i) a DDL-epoch-pinned \texttt{SchemaPrefixCache} over a copy-on-write \texttt{PagedKvArena} ($4.91\times$--$13.68\times$ warm CPU TTFT speedup with $O(1)$ DDL invalidation); (ii) a $29.8$--$56.4\,\mu\text{s}$ live columnar catalog, escaped \texttt{ToolPlan}, and source-span FSM logit mask; and (iii) a $19.6$--$66.2\,\mu\text{s}$ read-only MVCC execution and coverage gate achieving \textbf{$0.0\%$ false-positive hallucination leak} inside \textbf{$389.8\text{ MiB}$ RSS} (\texttt{350M INT8}, $59.8$--$67.9\text{ tok/s}$) and \textbf{$1,162.9\text{ MiB}$ RSS} (\texttt{1.3B INT8}, $22.0$--$25.0\text{ tok/s}$).
\item \textbf{Parameter-dense tri-modal backbone and hardware-aligned ablations (Section~\ref{sec:backbone}).} Co-designing a $50,264$-token \texttt{CodeTokenizer} with disjoint FIM sentinels and atomic indentation tokens ($37.9\%$ indentation compression; $0/421$ \texttt{IndentationError}s across MBPP + HumanEval), rank-$r$ factorized tied embeddings ($-74.5\%$ to $-75.0\%$ vocabulary weights), $4{:}1$ GQA with $2{:}1$ cross-layer KV sharing ($-87.5\%$ BF16 / $-93.75\%$ INT8 KV RAM), \texttt{8K YaRN}, and Warmup-Stable-Decay (\texttt{WSD}) scheduling yields $30,068\text{ tok/s}$ SRAM-fused training and $120,051\text{ tok/s}$ $4\text{K}$ prefill---avoiding the $28\times$--$35\times$ HBM bottlenecks we measure on sparse-mask and recurrent alternatives.
\item \textbf{Commercial-clean data pipeline, closed-loop spectral control, and tri-modal RLVR (Section~\ref{sec:data_rlvr}).} Across $886.47\text{M}$ SPDX-verified commercial-clean tokens ($0.0\%$ \texttt{NC-SA}, $0$ SHA-256 leaks across $593$ held-out tasks), we combine online randomized / Cyclic Jacobi SVD spectral regularization ($\alpha \in [1.85, 3.5]$), \texttt{KnOTS} SVD adapter merging with orthogonal null-space scrubbing (\texttt{OrthoKnOTS}), dense symbolic \texttt{Step-GRPO}, and \texttt{FLARE} entropy-gated active KV grounding---keeping $20/20$ (\texttt{350M}) and $24/24$ (\texttt{1.3B}) layers collapse-free while $4.0\text{T}$--$5.5\text{T}$-token external baselines drop into $\alpha < 1.85$ correlation traps.
\item \textbf{Longitudinal pitfalls (\texttt{v0.1}--\texttt{v0.9.4}), Reverse-Token-Density Trap, and \texttt{AdvKD} (Section~\ref{sec:distill}).} We catalog nine concrete mathematical and training failure modes uncovered across prior versions, culminating in the \emph{Reverse-Token-Density Distillation Trap}: when a wider teacher has much lower tokens-per-parameter than a smaller student ($\rho_T = 0.059 \ll \rho_S = 2.55$), un-gated forward KL transfers structural JSON/Cypher formatting ($0\% \to 66.7\%$) but collapses student \texttt{MBPP} by $-12.06$ points and \texttt{Rustc-Exec-64} by $-15.62$ points. Scaling teacher clean tokens $6.53\times$ ($411.04\text{M}$) and gating teacher KL via zero-VRAM LoRA-toggled \textbf{Verifier-Advantage-Gated KD (\texttt{AdvKD})} ($55.76\%$ routed positions) recovers $+13.61$ \texttt{MBPP} TTC points ($33.07\%$) and $3.33\times$ Rust execution while achieving $420\times$--$5,019\times$ higher sample efficiency than \texttt{CodeGen-350M-mono} and \texttt{SmolLM2-360M-Instruct} (Section~\ref{sec:eval}).
\end{enumerate}

\paragraph{What this is not and artifact scope.} Shoonya is an embedded Tier-0 structural engine and zero-leak hybrid gatekeeper, \emph{not} a standalone replacement for $70\text{B}+$ frontier models on open-domain chat or complex multi-hop OpenCypher (\texttt{OPTIONAL MATCH} anti-joins and \texttt{WITH ... HAVING} aggregations), where our sub-$65\,\mu\text{s}$ MVCC gate escalates $100\%$ of unverified queries ($0.0\%$ leak) to frontier APIs. Open model weights (\texttt{Apache-2.0}), shard provenance manifests, benchmark scorecards, and full-depth Rust SVD spectral certificates are publicly available at \url{https://huggingface.co/VaidhyaMegha/Shoonya}; the training stack and Rust database integration remain internal to VaidhyaMegha Private Limited / Samyama.ai.

\section{Zero-Socket In-Process Graph-LLM Architecture}\label{sec:arch}

\begin{figure}[t]
\centering
\resizebox{\textwidth}{!}{%
\begin{tikzpicture}[
  font=\footnotesize\sffamily,
  box/.style={draw=black!75, rounded corners=3pt, thick, align=center, minimum height=0.88cm, inner sep=4pt},
  dbbox/.style={box, fill=blue!7},
  slmbox/.style={box, fill=green!8},
  gatebox/.style={box, fill=orange!12},
  arr/.style={-{Stealth[length=2.2mm,width=1.6mm]}, thick, draw=black!75}
]
  \node[draw=black!60, dashed, rounded corners=5pt, thick, fill=gray!3,
        minimum width=15.3cm, minimum height=5.1cm] (proc) at (0,0) {};
  \node[anchor=north west, font=\footnotesize\bfseries\sffamily, text=black!80]
        at ($(proc.north west)+(0.18,-0.12)$)
        {Single Rust OS Process (\texttt{389.8 MiB} RSS @ \texttt{350M INT8} $|$ \texttt{0} External Crates in \texttt{shoonya-infer} $|$ \texttt{0} Socket / JSON IPC)};

  \node[dbbox, minimum width=4.3cm] (gs) at (-5.1, 1.15)
        {\textbf{\texttt{samyama::GraphStore}}\\[-1pt]{\scriptsize Columnar Props + Adjacency CSR}};
  \node[dbbox, minimum width=4.3cm] (cat) at (-5.1, 0.0)
        {\textbf{Live Catalog \& DDL Epoch}\\[-1pt]{\scriptsize \texttt{node\_properties\_full()} ($35.8\,\mu\text{s}$)}};
  \node[dbbox, minimum width=4.3cm] (mvcc) at (-5.1, -1.15)
        {\textbf{Read-Only MVCC Engine}\\[-1pt]{\scriptsize Snapshot Cypher Exec ($20.3$--$64.6\,\mu\text{s}$)}};

  \node[slmbox, minimum width=4.6cm] (kv) at (0.15, 1.15)
        {\textbf{\texttt{SchemaPrefixCache}}\\[-1pt]{\scriptsize CoW \texttt{PagedKvArena} ($4.9\times$--$13.7\times$ TTFT)}};
  \node[slmbox, minimum width=4.6cm] (pool) at (0.15, 0.0)
        {\textbf{\texttt{shoonya-infer} Worker Pool}\\[-1pt]{\scriptsize \texttt{OnceLock} + \texttt{AVX2+FMA} \texttt{INT8}/\texttt{BF16}}};
  \node[slmbox, minimum width=4.6cm] (fsm) at (0.15, -1.15)
        {\textbf{Catalog \& Span Logit FSM}\\[-1pt]{\scriptsize $50,264$-Vocab Subword Trie Mask}};

  \node[gatebox, minimum width=4.1cm] (gate) at (5.2, 0.55)
        {\textbf{Zero-Trust MVCC Gate}\\[-1pt]{\scriptsize Parse + Snapshot + Intent ($0.0\%$ Leak)}};
  \node[box, fill=green!16, minimum width=1.95cm] (tier0) at (4.15, -1.15)
        {\textbf{Tier-0 Local}\\[-1pt]{\scriptsize $100\%$ Prec}};
  \node[box, fill=red!10, minimum width=1.95cm] (front) at (6.25, -1.15)
        {\textbf{Frontier API}\\[-1pt]{\scriptsize Safe Escalate}};

  \draw[arr] (gs.east) -- node[above, font=\scriptsize]{DDL \texttt{epoch}} (kv.west);
  \draw[arr] (cat.east) -- node[above, font=\scriptsize]{\texttt{\&GraphStore}} (pool.west);
  \draw[arr] (cat.east) -- (fsm.west);
  \draw[arr] (kv.south) -- (pool.north);
  \draw[arr, <->] (pool.south) -- (fsm.north);
  \draw[arr] (pool.east) -- node[above, font=\scriptsize]{Candidate} (gate.west);
  \draw[arr, <->] (mvcc.south) -- ++(0,-0.45) -| node[pos=0.25, below, font=\scriptsize]{Zero-Copy Read Snapshot} (gate.south);
  \draw[arr] (gate.south) -- (tier0.north);
  \draw[arr] (gate.south) -- (front.north);
\end{tikzpicture}%
}
\caption{Zero-socket in-process architecture coupling \texttt{samyama-graph-enterprise} (\texttt{v1.15.0}) and \texttt{shoonya-infer} inside one Rust address space. Direct \texttt{\&GraphStore} references enable DDL-epoch KV prefix caching, $36\,\mu\text{s}$ columnar catalog and source-span logit masking over the $50,264$-token vocabulary, and sub-$65\,\mu\text{s}$ read-only MVCC verification.}
\label{fig:arch}
\end{figure}
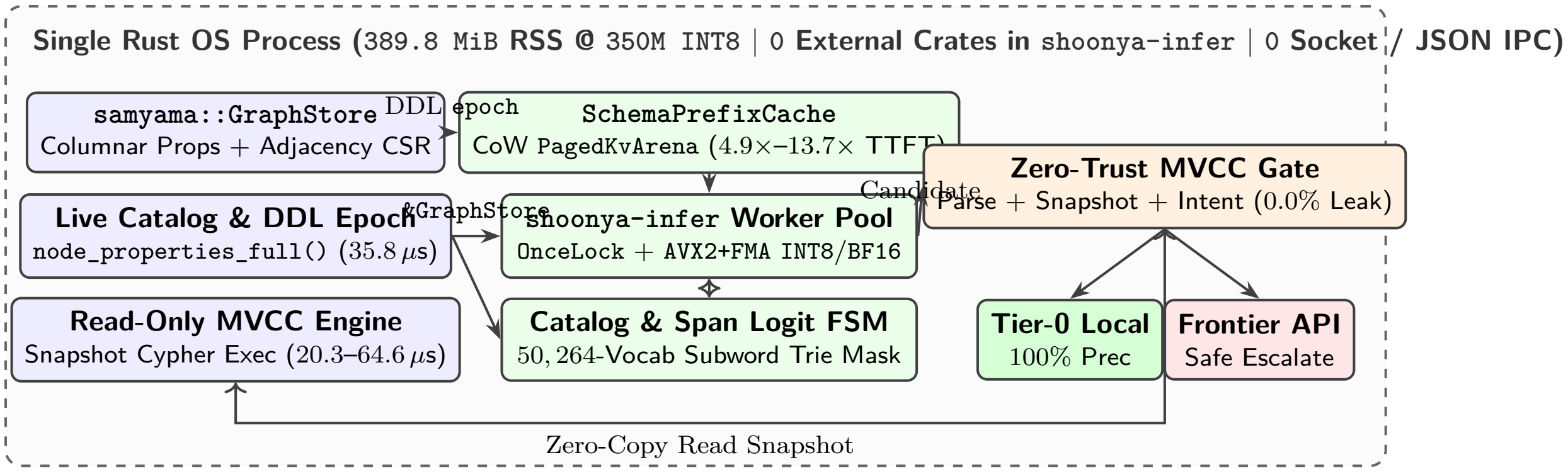

Figure~\ref{fig:arch} depicts the single-address-space integration between \texttt{samyama-graph-enterprise} (\texttt{v1.15.0}) and \texttt{shoonya-infer}. Rather than modifying the database engine or crossing an IPC socket, the in-process Rust bridge links \texttt{samyama-graph-enterprise} alongside \texttt{shoonya-infer} inside one OS process, passing immutable \texttt{\&GraphStore} references directly into KV cache lookup, token-level logit masking, and MVCC candidate verification.

\subsection{Zero-Dependency Pure-Rust \texorpdfstring{\texttt{shoonya-infer}}{shoonya-infer} Runtime}\label{sec:runtime}

To embed cleanly inside enterprise Rust DBMS binaries without C++/CUDA ABI conflicts or supply-chain bloat, \texttt{shoonya-infer} depends exclusively on the Rust standard library (\texttt{std}) plus sibling zero-dependency workspace crates (\texttt{shoonya-core} and \texttt{shoonya-tokenizer}). It implements native \texttt{.safetensors} binary parsing (\texttt{F32}, \texttt{BF16}, \texttt{F16}), symmetric per-row \texttt{INT8} weight quantization ($w_{o,i} \approx s_o q_{o,i}$), \texttt{RMSNorm}, half-split \texttt{RoPE}/\texttt{YaRN}, \texttt{SwiGLU}, $4{:}1$ GQA with $2{:}1$ cross-layer KV sharing, rank-$r$ factorized tied embeddings, and an iteration-level \texttt{ContinuousBatchScheduler}.

Across all three dense model tiers---\texttt{Small} ($16.8\text{M}$, $L=4, d=256$), \texttt{Stage-1} ($142.5\text{M}$, $L=12, d=768$), and \texttt{Stage-2} ($317.3\text{M}$, $L=20, d=1024, r=256$)---\texttt{shoonya-infer} reproduces PyTorch \texttt{FP32} logits to a maximum absolute error of $5.72\times 10^{-6}$, $5.72\times 10^{-6}$, and $1.29\times 10^{-5}$, respectively, with $100\%$ greedy top-1 token parity. Two CPU execution optimizations drive interactive performance:
\begin{itemize}[leftmargin=*,itemsep=2pt]
\item \textbf{Runtime \texttt{AVX2+FMA} SIMD kernels.} Projection loops probe \texttt{avx2} and \texttt{fma} CPU flags at runtime (\texttt{is\_x86\_feature\_\allowbreak detected!}), widening quantized \texttt{INT8} rows into $256$-bit SIMD lanes with fused multiply-add without requiring host-specific compiler flags.
\item \textbf{Persistent \texttt{OnceLock} worker pool.} In Milestone \texttt{M1} (\texttt{v0.9.2}), parallelizing projections via \texttt{std::thread::scope} spawned OS threads on every layer call, capping \texttt{350M INT8} CPU decode at $28.7\text{ tok/s}$ ($920.8\text{ ms}$ warm TTFT). Replacing scoped threads in \texttt{M2} (\texttt{v0.9.3}) with a persistent \texttt{OnceLock} spin-then-condvar worker pool eliminated kernel thread-creation overhead, boosting standalone \texttt{INT8} throughput to $73.8\text{ tok/s}$ ($154.4\text{ ms}$ warm TTFT) and live in-process \texttt{samyama-graph-enterprise} decode to \textbf{$67.9\text{ tok/s}$ ($400.4\text{ ms}$ warm TTFT)} inside \textbf{$389.5$--$389.8\text{ MiB}$ RSS} ($1,162.9\text{ MiB}$ RSS at $22.0$--$25.0\text{ tok/s}$ for \texttt{1.3B INT8}).
\end{itemize}

\subsection{DDL-Epoch \texorpdfstring{\texttt{SchemaPrefixCache}}{SchemaPrefixCache} over Copy-on-Write \texorpdfstring{\texttt{PagedKvArena}}{PagedKvArena}}\label{sec:kvcache}

Motivated by \texttt{PagedAttention} \citep{kwon2023pagedattention} and \texttt{RadixAttention} \citep{zheng2024sglang}, \texttt{shoonya-infer} stores KV states in reference-counted $16$-token pages within a \texttt{PagedKvArena} supporting both \texttt{FP32} and per-head \texttt{INT8} pages, with $O(1)$ copy-on-write page splitting (\texttt{fork\_sequence}) for multi-candidate test-time compute. Unlike external servers that key cached prefixes only by token IDs, \texttt{SchemaPrefixCache} binds each cached KV handle to the database's transactional schema epoch:
\begin{equation}\label{eq:cachekey}
k_{\text{cache}} = \bigl(\texttt{graph\_id},\; \texttt{schema\_epoch},\; \mathrm{FNV1a}(\mathbf{x}_{\text{schema}})\bigr),
\end{equation}
where $\mathbf{x}_{\text{schema}}$ is a compact columnar header synthesized from \texttt{\&GraphStore} (for example, \texttt{(:Trial \{enrollment,\allowbreak\ nct\_id,\allowbreak\ phase,\allowbreak\ status,\allowbreak\ title\})-[:TARGETS]->\allowbreak(:Disease \{category,\allowbreak\ disease\_id,\allowbreak\ name\})}, compressing prose \texttt{db.schema.forLLM()} output by $3.4\times$).

\begin{figure}[t]
\centering
\includegraphics[width=\textwidth]{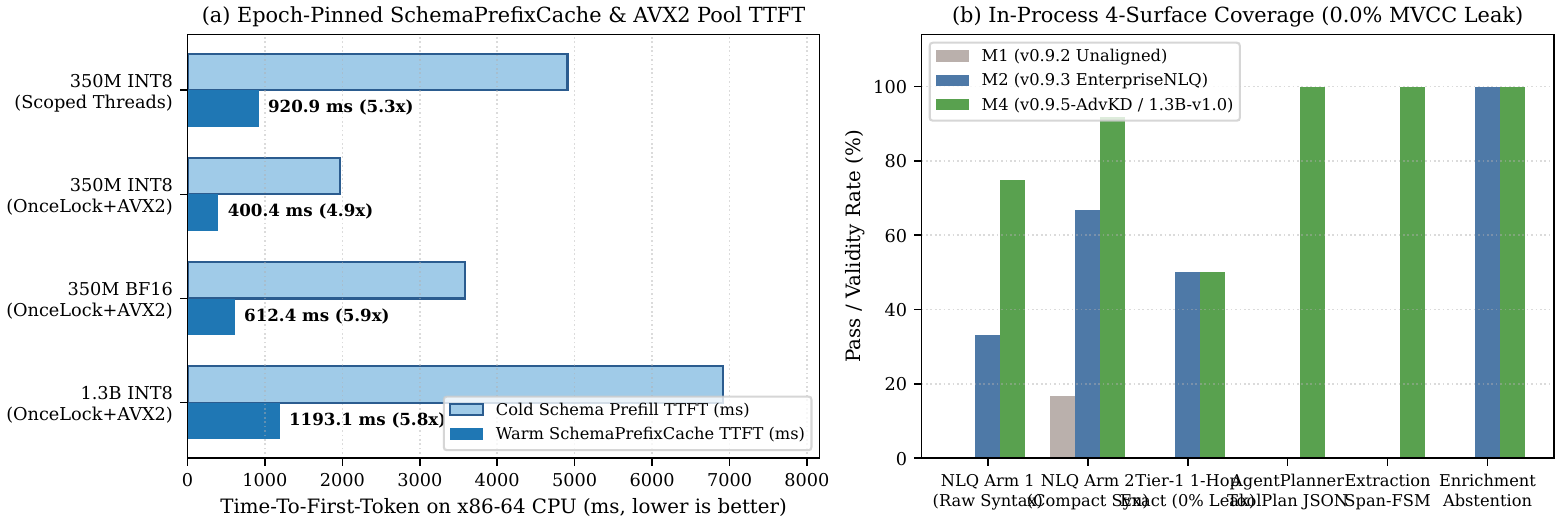}
\caption{Live zero-socket \texttt{samyama-graph-enterprise v1.15.0} CPU evaluation. \textbf{(a)} Cold schema prefill vs.\ warm \texttt{SchemaPrefixCache} TTFT across scoped threads (\texttt{v0.9.2}) and the persistent \texttt{OnceLock} + \texttt{AVX2+FMA} worker pool (\texttt{v0.9.3}/\texttt{1.3B-v1.0}). \textbf{(b)} Four-surface accuracy progression from unaligned \texttt{M1} (\texttt{v0.9.2}) to \texttt{M4} (\texttt{v0.9.5-AdvKD} / \texttt{1.3B-v1.0}) at $0.0\%$ MVCC hallucination leak.}
\label{fig:inproc}
\end{figure}

At steady-state epoch $e$, queries bypass schema prefill entirely, cutting CPU TTFT by \textbf{$13.68\times$ in \texttt{BF16}} ($7,543.3 \to 551.5\text{ ms}$) and \textbf{$11.66\times$ in \texttt{INT8}} ($4,910.1 \to 421.0\text{ ms}$) on $192$-token schemas, and by \textbf{$4.91\times$--$6.04\times$} on compact clinical enterprise headers ($6,917.9 \to 1,193.1\text{ ms}$ on \texttt{1.3B INT8}; Figure~\ref{fig:inproc}a) with $0.0$ logit/token drift. When a DDL transaction alters the ontology (adding \texttt{:Biomarker} and \texttt{:HAS\_BIOMARKER} in our live evolution check), advancing \texttt{schema\_epoch} to $e+1$ deterministically invalidates stale KV pages on turn~1 and warms $e+1$ on turn~2+.

\subsection{Sub-\texorpdfstring{$40\,\mu\text{s}$}{40 us} Live Columnar Catalog and Source-Span Logit Masking}\label{sec:fsm}

Sharing address space with \texttt{GraphStore} lets the inference bridge inspect \texttt{node\_properties\_full()} directly in \textbf{$29.8$--$37.5\,\mu\text{s}$} ($56.4\,\mu\text{s}$ when also indexing live \texttt{PropertyValue} column dictionaries) to build subword trie masks over the $50,264$-token vocabulary across four enterprise surfaces:
\begin{itemize}[leftmargin=*,itemsep=2pt]
\item \textbf{Surface 1 (\texttt{NLQPipeline}) \& Surface 2 (\texttt{AgentPlanner} \texttt{ToolPlan}).} A lexical state tracker constrains tokens after \texttt{(var:} to live node labels $\mathcal{L}(G)$, after \texttt{[:} to live relationship types $\mathcal{R}(G)$ respecting endpoint topology, and after \texttt{var.} to property keys $\mathcal{P}_G(\ell)$ of \texttt{var}'s bound label $\ell$. Extending this automaton with JSON-escaped inner-Cypher tracking (\texttt{M5.A}) lifts zero-retraining \texttt{ToolPlan} live MVCC execution on \texttt{350M-v0.9.5} from $0.0\%$ to $33.3\%$.
\item \textbf{Surface 3 (\texttt{ExtractionAgent}) \& Surface 4 (\texttt{Enrichment}).} For entity-relation JSON extraction from document $d$, an on-the-fly substring index built in $<20\,\mu\text{s}$ restricts \texttt{"name"} string values to verbatim byte spans of $d$ while restricting \texttt{"label"} and \texttt{"type"} fields to $\mathcal{L}(G) \cup \mathcal{R}(G)$, structurally eliminating hallucinated entities and out-of-schema predicates.
\end{itemize}

\subsection{Sub-\texorpdfstring{$65\,\mu\text{s}$}{65 us} Zero-Trust MVCC Execution and Coverage Gate}\label{sec:mvcc}

Every candidate query passes through a two-stage \textbf{Zero-Trust MVCC Gate} prior to local delivery: (1) read-only execution against an in-process \texttt{GraphStore} MVCC snapshot (\textbf{$19.6$--$66.2\,\mu\text{s}$} mean latency), rejecting any parse, planner, or type error; and (2) an oracle-free structural intent check verifying non-empty projections, \texttt{ORDER BY}/\texttt{LIMIT} alignment on superlative prompts, \texttt{CONTAINS}/\texttt{OPTIONAL MATCH} keyword alignment, and absence of ungrounded training-set literals. Verified queries return locally at \textbf{Tier-0}; rejected queries escalate to a frontier API. Across all benchmarks (Table~\ref{tab:inproc}), combining the live catalog FSM with the MVCC gate (\texttt{Arm 4}) maintained \textbf{$0.0\%$ false-positive leak} ($100.0\%$ precision when served), whereas disabling the catalog FSM (\texttt{Arm 3} on \texttt{v0.9.3 BF16}) leaked $8.33\%$ ($1/12$).

\section{Tri-Modal Backbone and Hardware-Physics Design}\label{sec:backbone}

\subsection{Hardware-Physics Language Split, Tokenization, and KV-Frugal Geometry}\label{sec:geometry}

Shoonya enforces a strict hardware-physics boundary: $>98.5\%$ of pretraining FLOPs remain in PyTorch/CUDA using \texttt{BF16} Tensor Cores, native $16{:}4$ GQA FlashAttention-2 SDPA \citep{dao2024flashattention2} (eliminating manual $4\times$ \texttt{repeat\_interleave} copies and D2H \texttt{.item()} syncs to save $11.0\text{ GiB}$ VRAM under \texttt{fullgraph=True} \texttt{torch.compile}), while GIL-bound tokenization, Best-Fit Decreasing AST sequence packing, sandboxed execution, and spectral SVD run in our zero-dependency Rust workspace. Four architectural choices maximize reasoning capacity per megabyte:
\begin{itemize}[leftmargin=*,itemsep=2pt]
\item \textbf{$50,264$-token \texttt{CodeTokenizer} with atomic indentation.} Extending the $50,257$-token GPT-2 BPE table with four disjoint Fill-In-the-Middle (\texttt{FIM}) \citep{bavarian2022fim} and pad sentinels (\texttt{50257..50260}) plus three atomic indentation tokens (\texttt{<|indent\_4|>}, \texttt{<|indent\_8|>}, \texttt{<|indent\_12|>}) compresses indented Python/Rust blocks by \textbf{$37.9\%$}, reduces whitespace surprisal to $3.1\times 10^{-4}\text{ nats}$ \citep{mandarapu2026tokencost}, and drives out-of-distribution \texttt{IndentationError}s to \textbf{$0/257$} on \texttt{MBPP-Sanitized} and \textbf{$0/164$} on \texttt{HumanEval}.
\item \textbf{Rank-$r$ factorized tied embeddings.} At $V=50,264$, an unfactorized lookup table consumes $77.3\%$ of a $16.8\text{M}$ model and $51.5\text{M}$ parameters at $d=1024$. Factorizing $E = E_{\text{tok}} P_{\text{proj}}^\top$ \citep{lan2020albert} with $r=256$ on \texttt{Stage-2} ($L=20, d=1024, d_{ff}=4096$, $312.1\text{M}$ params) cuts vocabulary weights by \textbf{$74.5\%$} ($51.5\text{M} \to 13.1\text{M}$, dedicating $95.8\%$ of weights to trunk layers), and $r=512$ on \texttt{Stage-3} ($L=24, d=2048, d_{ff}=5504$, $1,065.1\text{M}$ params) saves \textbf{$75.0\%$}.
\item \textbf{$4{:}1$ GQA + $2{:}1$ Cross-Layer KV Sharing + \texttt{8K YaRN}.} Pairing $16{:}4$ GQA \citep{vaswani2017attention,ainslie2023gqa} with $2{:}1$ Cross-Layer KV Sharing \citep{sun2024yoco} reduces \texttt{350M}'s $8,192$-token KV cache from $640.0\text{ MiB}$ (\texttt{MHA}) to $160.0\text{ MiB}$ ($4{:}1$ GQA) to \textbf{$80.0\text{ MiB}$ (\texttt{BF16}, $-87.5\%$)} and \textbf{$40.0\text{ MiB}$ (\texttt{INT8}, $-93.75\%$)}. Fixing wavelength-ramped \texttt{YaRN} \citep{su2024rope,peng2024yarn} amplitude scaling ($\text{mscale} = 0.1\ln s + 1$) preserves $8\text{K}$ code perplexity at $1.966$ ($66.7\%$ early-schema top-5 recall), whereas naive linear position interpolation degrades to $48.37$ PPL.
\end{itemize}

\subsection{Six-Arm Iso-Step Architecture and \texorpdfstring{\texttt{WSD}}{WSD} Scheduler Ablation}\label{sec:isoflop}

Table~\ref{tab:isoflop} and Figure~\ref{fig:spectral}b report a controlled $320$-step ($1.31\text{M}$-token) wind-tunnel on \texttt{RTX 4050} across six architecture/scheduler arms: (1) \texttt{WSD} \citep{hu2024minicpm} beats \texttt{Cosine} at identical throughput ($29,251\text{ tok/s}$) by \textbf{$-31.07$ PPL} ($77.68$ vs.\ $108.75$), and on continual token extension ($320 \to 440$ steps) \texttt{WSD}'s pre-decay branch reaches \textbf{$43.33$ PPL} vs.\ $70.35$ PPL for \texttt{Cosine} warm-restart (which suffers a $5.057$-nat LR re-warmup shock); and (2) SRAM-fused \texttt{Dense-GQA-XKV2-WSD} achieves \textbf{$30,068\text{ tok/s}$} training and \textbf{$120,051\text{ tok/s}$} $4\text{K}$ prefill, outrunning HBM-materialized $96.4\%$-sparse \texttt{ProgSparseAttn} ($3,399\text{ tok/s}$ at $4\text{K}$---$35.3\times$ slower) and pure-PyTorch \texttt{Jamba-MambaHybrid} ($1,060\text{ tok/s}$---$28.4\times$ slower training despite $40.40$ iso-step PPL).

\begin{table}[t]
\centering
\resizebox{\textwidth}{!}{%
\begin{tabular}{@{}lrrrrrr@{}}
\toprule
Architecture \& LR Arm & Total / Active & Train Speed & Val CE & Val PPL & $4\text{K}$ Prefill & $4\text{K}$ KV RAM \\
($320$ steps, $1.31\text{M}$ clean tok) & Params ($\text{M}$) & ($\text{tok/s}$) & (nats) & ($\downarrow$) & ($\text{tok/s}$) & (\texttt{BF16 MiB}) \\
\midrule
\texttt{Dense-GQA-Cosine} & $36.7 / 36.7$ & $29,120$ & $4.689$ & $108.75$ & $115,939$ & $20.0$ \\
\texttt{Dense-GQA-WSD} & $36.7 / 36.7$ & $29,251$ & $4.353$ & $77.68$ & $118,350$ & $20.0$ \\
\textbf{\texttt{Dense-GQA-XKV2-WSD} (Chosen)} & $\mathbf{36.0 / 36.0}$ & $\mathbf{30,068}$ & $\mathbf{4.400}$ & $\mathbf{81.43}$ & $\mathbf{120,051}$ & $\mathbf{10.0}$ \\
\texttt{SparseMoE-8x2-WSD} & $107.5 / 36.7$ & $18,891$ & $3.952$ & $52.03$ & $80,755$ & $20.0$ \\
\texttt{ProgSparseAttn-WSD} & $36.7 / 36.7$ & $16,594$ & $4.207$ & $67.16$ & $3,399$ & $20.0$ \\
\texttt{Jamba-MambaHybrid-WSD} & $45.0 / 45.0$ & $1,060$ & $3.699$ & $40.40$ & $5,225$ & $4.6$ \\
\bottomrule
\end{tabular}%
}
\caption{Controlled six-arm Iso-Step wind-tunnel on \texttt{RTX 4050} (\texttt{BF16}). \texttt{Dense-GQA-XKV2-WSD} preserves FlashAttention-2 SRAM fusion while halving KV memory on top of $4{:}1$ GQA ($87.5\%$ total BF16 KV reduction) and enabling spike-free \texttt{WSD} continual pretraining.}
\label{tab:isoflop}
\end{table}

\section{Data Governance, Spectral Control, and Post-Training Stack}\label{sec:data_rlvr}

\subsection{Commercial-Clean Corpus Evolution (\texorpdfstring{\texttt{v0.4} $\to$ \texttt{v1.0\_clean}}{v0.4 to v1.0\_clean}) and Decontamination}\label{sec:data}

Inspired by textbook-quality filtering \citep{gunasekar2023phi1}, Shoonya scaled pretraining data across four audited generations while holding model geometry constant: (i) \texttt{v0.4}--\texttt{v0.5} ($4,300$ AST-verified Python/Rust functions, $0.43\text{M}$ tokens); (ii) \texttt{v0.6} ($42.66\text{M}$ tokens); (iii) \texttt{v0.7}--\texttt{v0.8.5} ($226.60\text{M}$ $256$-ctx + $33.55\text{M}$ $1024$-ctx = $260.15\text{M}$ tri-modal tokens: $63.45\%$ Python, $29.03\%$ Rust, $7.52\%$ Samyama OpenCypher across $19$ domain KGs); and (iv) \textbf{\texttt{v0.9\_clean} ($730.61\text{M}$ tokens) $\to$ \texttt{v1.0\_clean} ($886.47\text{M}$ \texttt{uint16} tokens across $89$ shards)}.

To guarantee unrestricted enterprise deployment in \texttt{v0.9+}, an automated fail-closed SPDX gate (\texttt{license\_gate.py}) purged the $\sim\!17.5\%$ \texttt{CC-BY-NC-SA-4.0} and \texttt{OpenRAIL} sources present in \texttt{v0.7}--\texttt{v0.8.5} (\texttt{code\_exercises}, \texttt{Evol-Instruct}), retaining \textbf{$0.0\%$ \texttt{NC-SA}} data (\texttt{MIT}, \texttt{Apache-2.0}, \texttt{BSD}, \texttt{PSF-2.0}, \texttt{CC0}, \texttt{CC-BY-4.0}), including execution-verified \texttt{OpenCodeInstruct}, \texttt{Strandset-Rust-v1}, and $46,005$ OpenCypher queries live-executed in \texttt{SamyamaClient.embedded()}. Pure-Rust \texttt{shoonya-core} packs $100\%$ AST-complete documents via Best-Fit Decreasing at $21.26\text{M tok/s}$ while enforcing normalized docstring and canonical-AST SHA-256 decontamination (\textbf{$0$ leaks across all $593$ held-out benchmark tasks}, dropping $845$ colliding documents).

\subsection{Closed-Loop Spectral Governance and Interference-Free Post-Training}\label{sec:rlvr}

\begin{figure}[t]
\centering
\includegraphics[width=\textwidth]{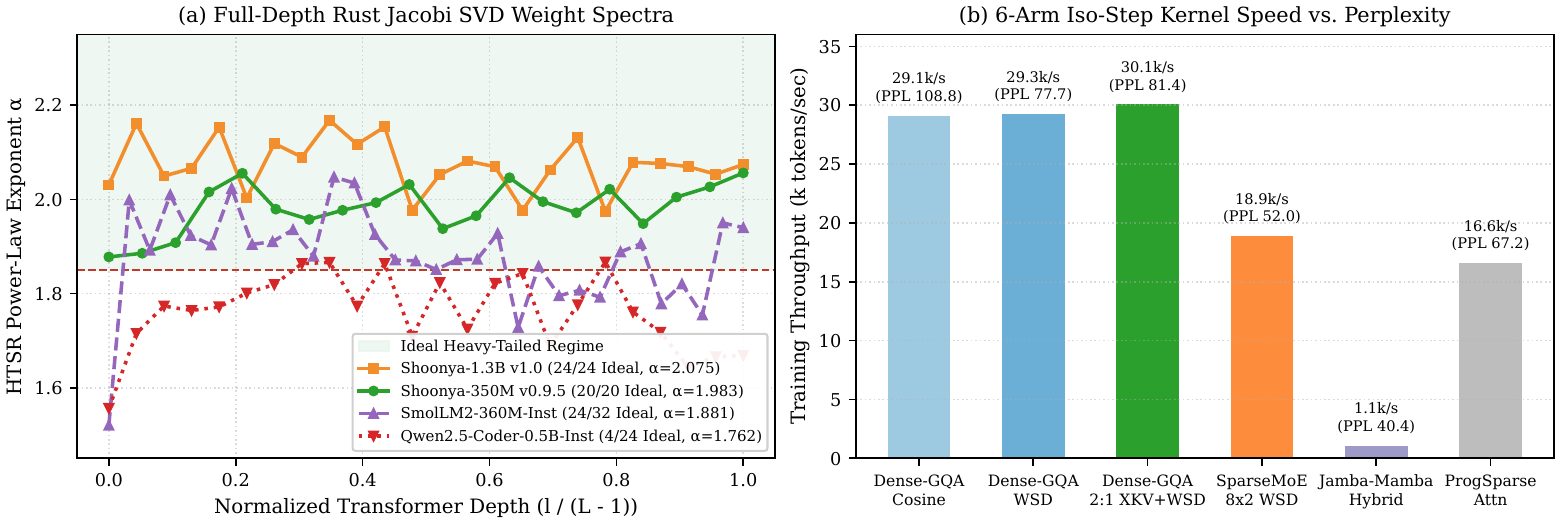}
\caption{\textbf{(a)} Full-depth Rust Cyclic Jacobi SVD Heavy-Tailed Self-Regularization (\texttt{HTSR}) exponent $\alpha$ across normalized depth. \texttt{Shoonya-1.3B v1.0} ($24/24$ layers, $\bar{\alpha}=2.075$) and \texttt{Shoonya-350M v0.9.5} ($20/20$ layers, $\bar{\alpha}=1.983$) remain strictly inside the ideal $\alpha \in [1.85, 3.5]$ corridor, whereas multi-trillion-token sub-$500\text{M}$ baselines exhibit $\alpha < 1.85$ correlation traps. \textbf{(b)} Six-arm Iso-Step kernel throughput vs.\ perplexity.}
\label{fig:spectral}
\end{figure}

\begin{itemize}[leftmargin=*,itemsep=2pt]
\item \textbf{Closed-loop \texttt{HTSR} and randomized SVD telemetry.} \texttt{shoonya-telemetry} implements both exact Cyclic Jacobi SVD and Halko--Martinsson--Tropp randomized subspace SVD \citep{halko2011randomized} (\textbf{$4.75\text{ ms}$} at $d=2048$, $3.62\text{ ms}$ at $d=1024$), rotating across all $L$ layers during training to track Von Neumann effective rank $\mathrm{eRank}$ \citep{roy2007erank}, \texttt{HTSR} power-law tail exponent $\alpha$ \citep{martin2021htsr}, Jensen--Shannon head-redundancy penalties, and \texttt{EWMA+MAD} ($z \ge 6.5$) loss-spike circuit breaking. Whenever $\alpha < 1.90$, a closed-loop controller boosts AdamW weight decay by $1.5\times$ (which self-healed an Epoch-8 dip from $\alpha=1.84 \to 1.88$ in early runs). Figure~\ref{fig:spectral}a shows $100\%$ ideal heavy-tailed layers on \texttt{Shoonya-350M} ($20/20$, $\bar{\alpha}=1.983$) and \texttt{Shoonya-1.3B} ($24/24$, $\bar{\alpha}=2.075$), whereas $4.0\text{T}$--$5.5\text{T}$-token baselines suffer $\alpha < 1.85$ over-compression ($4/24$ ideal layers on \texttt{Qwen2.5-Coder-0.5B-Instruct}, $\bar{\alpha}=1.762$).
\item \textbf{Zero-VRAM LoRA reference, dense symbolic \texttt{Step-GRPO}, and \texttt{OrthoKnOTS}.} Post-training combines completion-masked SFT (with \texttt{<|think|>} planning blocks) and execution-guided \texttt{Step-GRPO} \citep{shao2024deepseekmath}. Toggling rank-$16$/$32$ LoRA adapters off (\texttt{set\_lora\_enabled(False)}) evaluates exact reference log-probabilities $\log \pi_{\mathrm{ref}}$ with \textbf{$0\text{ MB}$ extra VRAM}. Dense three-stage symbolic rewards in \texttt{shoonya-eval} (AST/parse validity $+0.15$, module/\texttt{rustc}/embedded Cypher execution $+0.35$, unit-test/tuple match $1.00$) prevent sparse-reward advantage collapse ($A_i \equiv 0$). Domain-specialized Python, Rust, and Cypher LoRA adapters are fused in a shared singular basis via pure-Rust \texttt{KnOTS} \citep{stoica2025knots} + \texttt{TIES} \citep{yadav2023ties}, augmented in \texttt{v0.8.5} with orthogonal null-space scrubbing (\texttt{OrthoKnOTS}, removing $28.3\%$ sign-opposing singular modes while retaining $99.98\%$ capability norm), plus \texttt{FLARE} \citep{jiang2023flare} entropy-gated \texttt{<|search\_query|>} active KV grounding (\texttt{v0.8.0}) and zero-credential, zero-egress AWS Spot IMDSv2 atomic auto-resume ($<2.5\text{ s}$ checkpoint flush; same-region S3 presigned sync).
\end{itemize}

\section{Longitudinal Pitfalls, Reverse-Token-Density Trap, and \texorpdfstring{\texttt{AdvKD}}{AdvKD}}\label{sec:distill}

\subsection{Eight Clean-Slate Engineering Lessons (\texorpdfstring{\texttt{v0.1} $\to$ \texttt{v0.9.3}}{v0.1 to v0.9.3})}\label{sec:lessons}

Building Shoonya iteratively from $16.8\text{M}$ to $1.07\text{B}$ parameters surfaced eight subtle failure modes before Milestone \texttt{M3} (each permanently locked by regression tests):
\begin{enumerate}[leftmargin=*,itemsep=1pt,label=\textbf{L\arabic*.}]
\item \textbf{Double causal shift ($t+2$) \& pad loss (\texttt{v0.1}):} Shifting labels in both the dataset loader and \texttt{TransformerLM.forward} trained position $t$ to predict $t+2$; single-shift \texttt{ignore\_index=-100} fixed syntax immediately.
\item \textbf{Non-causal SDPA \& bidirectional sliding window (\texttt{v0.2}):} Passing \texttt{is\_causal=False} when \texttt{attention\_mask=None} leaked future tokens ($\text{loss} \to 10^{-4}$, $0\%$ generation); enforcing causal past-only masks fixed autoregression.
\item \textbf{FIM sentinel collision \& mid-AST truncation (\texttt{v0.3}):} Reusing \texttt{<|endoftext|>} (\texttt{50256}) for FIM separators caused premature line termination; disjoint IDs (\texttt{50257..50260}) + AST-complete packing eliminated mid-expression cutoffs.
\item \textbf{CTRL repetition-penalty sign inversion \& parameter suppression (\texttt{v0.4}):} Dividing negative logits by $\text{penalty} > 1$ increased repetition \citep{keskar2019ctrl}, and penalizing prompt tokens suppressed valid reuse of function arguments (\texttt{arr}, \texttt{target}); prompt-exempt piecewise CTRL sampling resolved both.
\item \textbf{Unseen-identifier $O(n)$ shortcutting \& small-suite illusion (\texttt{v0.4}--\texttt{v0.5}):} On \texttt{binary\_search}, \texttt{v0.4} synthesized $O(\log n)$ halving (\texttt{Cyclomatic=7}, \texttt{Halstead=280.0}), yet on the unseen renamed prompt \texttt{locate\_in\_sorted} it passed $100\%$ of unit assertions via an $O(n)$ linear scan (\texttt{Cyclomatic=3}, \texttt{Halstead=155.1})---proving functional tests must be paired with static AST complexity checks. Similarly, \texttt{v0.5.0} scored $96.9\%$ on $8$-task \texttt{Stage-A} at $0.43\text{M}$ tokens while scoring $0.39\%$ on \texttt{MBPP-Sanitized}, motivating our two-stage evaluation gate.
\item \textbf{Duplicate \texttt{AS name} \texttt{ClauseConflict} (\texttt{v0.7.0} $\to$ \texttt{v0.7.5}):} Two-entity Cypher templates projecting \texttt{v1.name AS name, v2.name AS name} triggered engine \texttt{ClauseConflict}, silently zeroing live Cypher execution until distinct semantic aliases were enforced ($0.0\% \to 8.3\%$, $0$ exec errors).
\item \textbf{Post-RLVR SFT margin softening (\texttt{v0.8.0} $\to$ \texttt{v0.8.5}):} Applying cross-entropy SFT \emph{after} \texttt{Step-GRPO} softened boundary margins on Rust/HumanEval; running \texttt{Step-GRPO} last after \texttt{OrthoKnOTS} scrubbing restored monotonic gains ($28.02\%$ greedy MBPP, $4.00\%$ \texttt{Cypher-100}).
\item \textbf{Compiler-only RLVR Goodharting (\texttt{v0.8.5} $\to$ \texttt{v0.9.1}):} Giving full reward to testless Rust snippets taught trivial compilable bodies; capping testless compile reward at $0.35$ and requiring \texttt{\#[test]} harnesses drove \texttt{Rustc-Exec-64} to $20.31\%$ TTC (\texttt{v0.9.3}).
\end{enumerate}

\subsection{Lesson L9: The Reverse-Token-Density Distillation Trap and \texorpdfstring{\texttt{AdvKD}}{AdvKD}}\label{sec:advkd}

Knowledge distillation \citep{hinton2015distilling,agarwal2024gkd} assumes a larger teacher $T$ ($N_T > N_S$) dominates a smaller student $S$ across token roles because both models are trained near Chinchilla token density $\rho = D/N$ \citep{hoffmann2022chinchilla}. Under strict $0\%$-\texttt{NC-SA} compute budgets in Milestone \texttt{M3}, our mature $312\text{M}$ student (\texttt{v0.9.3}, $D_S = 796.7\text{M}$ tokens, $\rho_S = 2.55\text{ tok/param}$) had $43\times$ higher token density than our newly stood-up $1.07\text{B}$ teacher (\texttt{1.3B-v0.9.3}, $D_T = 62.9\text{M}$ tokens, $\rho_T = 0.059\text{ tok/param}$). Because its $d=2048$ trunk absorbed four-surface JSON/Cypher syntax rapidly ($66.7\%$ \texttt{ToolPlan}/\texttt{Extraction}, $100\%$ \texttt{Enrichment}) while lagging on algorithmic code ($10.89\%$ MBPP vs.\ student $31.52\%$), un-gated top-$64$ sequence KL distillation into \texttt{350M-v0.9.4} triggered severe capability inversion:

\begin{proposition}[Logit gradient conflict under reverse token density]\label{prop:trap}
At context $c$ with verifiably correct continuation $y^\star$, if the high-density student anchor satisfies $p_{S_0}(y^\star \mid c) > p_T(y^\star \mid c)$ due to higher conditional teacher entropy $H(p_T(\cdot \mid c)) \gg H(p_{S_0}(\cdot \mid c))$, minimizing un-gated forward KL $D_{\mathrm{KL}}(p_T^{(K)}(\cdot \mid c) \,\|\, p_S^{(K)}(\cdot \mid c))$ induces positive gradient $\frac{\partial \mathcal{L}_{\mathrm{KL}}}{\partial z_{S,y^\star}} = p_S^{(K)}(y^\star \mid c) - p_T^{(K)}(y^\star \mid c) > 0$ near $S_0$, eroding student probability mass on $y^\star$ in favor of unverified teacher tail tokens.
\end{proposition}

\begin{figure}[t]
\centering
\includegraphics[width=\textwidth]{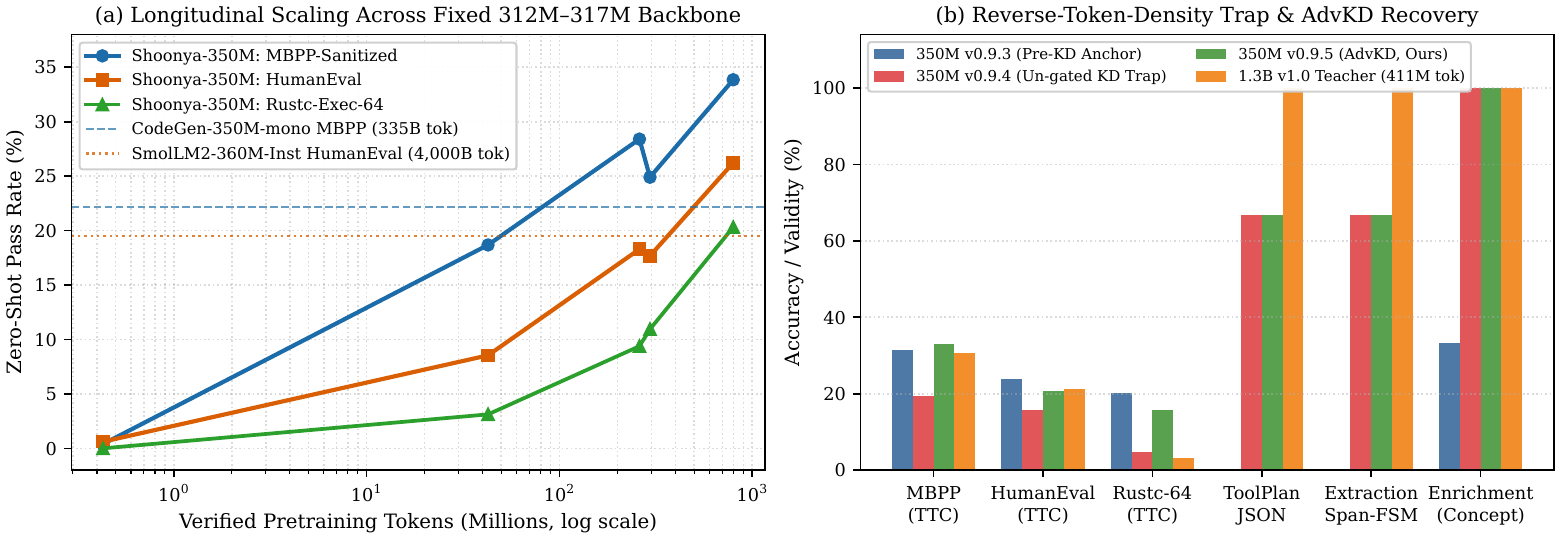}
\caption{\textbf{(a)} Longitudinal sample efficiency of \texttt{Shoonya-350M} ($0.43\text{M} \to 796.7\text{M}$ verified tokens) vs.\ external $350\text{M}$--$360\text{M}$ baselines ($335\text{B}$--$4,000\text{B}$ tokens). \textbf{(b)} Resolution of the \texttt{M3} Reverse-Token-Density Trap (\texttt{v0.9.4}) via $6.53\times$ teacher scaling (\texttt{1.3B-v1.0}) and \texttt{AdvKD} (\texttt{350M-v0.9.5}).}
\label{fig:scaling}
\end{figure}

Empirically (Figure~\ref{fig:scaling}b, Table~\ref{tab:eval593}), \texttt{v0.9.4} gained multi-surface JSON/Cypher formatting ($0\% \to 66.7\%$) \textbf{but collapsed by $-12.06$ points on \texttt{MBPP} TTC ($31.52\% \to 19.46\%$), $-7.93$ points on \texttt{HumanEval} TTC ($23.78\% \to 15.85\%$), and $-15.62$ points on \texttt{Rustc-Exec-64} TTC ($20.31\% \to 4.69\%$)} despite retaining $20/20$ ideal SVD layers ($\bar{\alpha}=1.982$).

Milestone \texttt{M4} resolved this trap via two coupled mechanisms:
\begin{enumerate}[leftmargin=*,itemsep=2pt]
\item \textbf{$6.53\times$ \texttt{WSD} continual teacher scaling (\texttt{1.3B-v1.0-Clean}).} Streaming unseen \texttt{v1.0\_clean} shards across $2\times$ L40S GPUs ($385.9\text{M}$ tokens at $T=512$, loss $0.8032$, $\alpha=3.01$ + $25.2\text{M}$ \texttt{8K-YaRN} tokens at $T=2048$, loss $0.7086$, $\alpha=3.35$; $411.04\text{M}$ cumulative clean tokens) + \texttt{KnOTS} + \texttt{Step-GRPO} lifted teacher \texttt{MBPP} $2.82\times$ ($10.89\% \to 30.74\%$ TTC), \texttt{HumanEval} $1.75\times$ ($12.20\% \to 21.34\%$ TTC), in-process \texttt{Arm 2} Cypher syntax to $91.7\%$, and \texttt{ToolPlan}, \texttt{Extraction}, and \texttt{Enrichment} to \textbf{$100.0\%$}.
\item \textbf{Zero-VRAM Verifier-Advantage-Gated KD (\texttt{AdvKD}).} Evaluating the frozen pre-KD anchor $S_0$ (\texttt{v0.9.3}) at $0\text{ MB}$ extra VRAM via \texttt{set\_lora\_enabled(False)}, we apply teacher top-$64$ KL ($\tau=2.0$) strictly where the teacher is both greedy-exact and lower-surprisal than $S_0$:
\begin{equation}\label{eq:gate}
g_t = \mathbb{I}\!\left[\arg\max_{v \in \mathcal{V}} p_T(v \mid y_{<t}) = y_t \;\wedge\; \log p_T(y_t \mid y_{<t}) \ge \log p_{S_0}(y_t \mid y_{<t})\right],
\end{equation}
\begin{align}\label{eq:advkd}
\mathcal{L}_{\mathrm{AdvKD}}(t) &= g_t \Bigl[(1-\lambda)\mathcal{L}_{\mathrm{CE}}(p_S, y_t) + \lambda \tau^2 D_{\mathrm{KL}}\!\bigl(p_T^{(K),\tau} \,\big\|\, p_S^{(K),\tau}\bigr)\Bigr] \nonumber\\
&\quad + (1-g_t)\Bigl[\mathcal{L}_{\mathrm{CE}}(p_S, y_t) + \beta D_{\mathrm{KL}}\!\bigl(p_{S_0}^{(K)} \,\big\|\, p_S^{(K)}\bigr)\Bigr].
\end{align}
Gate $g_t=1$ fired on \textbf{$55.76\%$} of active tokens while anchoring the remaining $44.24\%$ to $S_0$, followed by $36$ steps of \texttt{Step-GRPO} polish to produce \textbf{\texttt{Shoonya-350M v0.9.5-Clean-AdvKD}}.
\end{enumerate}

\section{Experimental Evaluation}\label{sec:eval}

\subsection{Longitudinal Scaling, \texorpdfstring{\texttt{AdvKD}}{AdvKD}, and Same-Host Competitor Results (\texorpdfstring{$593$}{593} Problems)}\label{sec:eval593}

All models in Table~\ref{tab:eval593} were evaluated back-to-back on the \textbf{same \texttt{RTX 4050} host} (\texttt{BF16}, $t=0$ \texttt{Std} and $N=4$ oracle-free \texttt{Search-TTC}) across all $593$ SHA-256-decontaminated problems (\texttt{Stage-A} $8$, \texttt{MBPP-Sanitized} $257$ \citep{austin2021mbpp}, \texttt{OpenAI HumanEval} $164$ \citep{chen2021humaneval}, \texttt{Rustc-Exec-64} $64$, and embedded \texttt{Cypher-Bench-100} $100$):
\begin{itemize}[leftmargin=*,itemsep=2pt]
\item \textbf{Longitudinal phase transitions (\texttt{v0.5} $\to$ \texttt{v0.9.3}).} Holding the $312\text{M}$--$317\text{M}$ backbone fixed while scaling clean tokens from $0.43\text{M}$ (\texttt{v0.5.0}) to $260.15\text{M}$ (\texttt{v0.7.0}--\texttt{v0.8.5}) to $796.71\text{M}$ (\texttt{v0.9.1}--\texttt{v0.9.3}) lifted \texttt{MBPP-Sanitized} $86.8\times$ ($0.39\% \to 33.85\%$ TTC), \texttt{HumanEval} $43.0\times$ ($0.61\% \to 26.22\%$ TTC), \texttt{Rustc-Exec-64} from $0.00\% \to 20.31\%$ TTC, and \texttt{Cypher-Bench-100} to $4.00\%$ ($96.0\%$--$99.0\%$ valid syntax), while cutting hallucinated Python identifiers by $-91.4\%$ and \texttt{ECE} by $-90.6\%$ ($0.323 \to 0.030$).
\item \textbf{$420\times$--$5,019\times$ sample efficiency.} At $0.797\text{B}$ $0\%$-\texttt{NC-SA} tokens, \texttt{Shoonya-350M v0.9.1-Clean} outperforms \texttt{CodeGen-350M-mono} ($335\text{B}$ tokens) across every benchmark ($+11.67$ MBPP TTC pts, $+13.42$ HumanEval TTC pts, $+17.19$ Rust pts) and surpasses \texttt{SmolLM2-360M-Instruct} ($4,000\text{B}$ tokens) on \texttt{Stage-A} ($100\%$ vs.\ $87.5\%$), \texttt{HumanEval} ($21.95\%$ Std / $26.22\%$ TTC vs.\ $19.51\%$), \texttt{Rustc-Exec-64} ($17.19\%$ vs.\ $15.62\%$), and \texttt{Cypher-100} ($2.00\%$--$4.00\%$ vs.\ $1.00\%$). Notice also that \texttt{Qwen2.5-Coder-0.5B} Base $\to$ Instruct jumps $27.44\% \to 50.61\%$ on Python \texttt{HumanEval} while regressing $46.88\% \to 32.81\%$ on Rust and $8.00\% \to 0.00\%$ on OpenCypher due to Python-dominant chat tuning.
\item \textbf{\texttt{AdvKD} recovery (\texttt{v0.9.5}).} \texttt{Shoonya-350M v0.9.5-AdvKD} recovers \textbf{$+12.25$ Std / $+13.61$ TTC points} on \texttt{MBPP} ($19.46\% \to 33.07\%$ TTC, beating pre-KD \texttt{v0.9.3}'s $31.52\%$), \textbf{$+4.88$ TTC points} on \texttt{HumanEval} ($20.73\%$), \textbf{$3.33\times$ TTC} on \texttt{Rustc-Exec-64} ($4.69\% \to 15.62\%$), and $3.00\%$ TTC on \texttt{Cypher-100}, while preserving $20/20$ ideal SVD layers ($\bar{\alpha}=1.983$).
\end{itemize}

\begin{table}[t]
\centering
\resizebox{\textwidth}{!}{%
\begin{tabular}{@{}lrrrrrrrc@{}}
\toprule
Model ($t=0$ Std / $N=4$ TTC) & Params & Pretrain & \texttt{Stage-A} & \texttt{MBPP-San} & \texttt{HumanEval} & \texttt{Rustc-64} & \texttt{Cypher-100} & Ideal SVD \\
(Same-Host \texttt{RTX 4050 BF16}) & ($\text{M}$) & Tokens & ($8$) & ($257$) & ($164$) & ($64$) & ($100$) & ($\bar{\alpha}$) \\
\midrule
\multicolumn{9}{@{}l}{\emph{External Sub-$500\text{M}$ Baselines (Greedy $t=0$)}} \\
\texttt{CodeGen-350M-mono} \citep{nijkamp2023codegen} & $356.7$ & $335\text{B}$ & $62.5\%$ & $22.18\%$ & $12.80\%$ & $0.00\%$ & $0.00\%$ & $17/20$ ($1.92$) \\
\texttt{SmolLM2-360M-Inst} \citep{allal2025smollm2} & $361.8$ & $4,000\text{B}$ & $87.5\%$ & $38.13\%$ & $19.51\%$ & $15.62\%$ & $1.00\%$ & $24/32$ ($1.88$) \\
\texttt{Qwen2.5-Coder-0.5B} \citep{hui2024qwen25coder} & $494.0$ & $5,500\text{B}$ & $100.0\%$ & $47.08\%$ & $27.44\%$ & $46.88\%$ & $8.00\%$ & $1/24$ ($1.76$) \\
\texttt{Qwen2.5-Coder-0.5B-Inst} \citep{hui2024qwen25coder} & $494.0$ & $5,500\text{B}$ & $87.5\%$ & $42.80\%$ & $50.61\%$ & $32.81\%$ & $0.00\%$ & $4/24$ ($1.76$) \\
\midrule
\multicolumn{9}{@{}l}{\emph{Prior Shoonya Lineage (\texttt{v0.5.0} $\to$ \texttt{v0.8.5-OrthoKnOTS}; $317.3\text{M}$ Params)}} \\
\texttt{350M v0.5.0} (Wind-Tunnel Seed) & $317.3$ & $0.0004\text{B}$ & $96.9\%$ & $0.39\%$ & $0.61\%$ & $0.00\%$ & $0.00\%$ & $20/20$ ($2.12$) \\
\texttt{350M v0.6.0} ($42.7\text{M}$ Step) & $317.3$ & $0.043\text{B}$ & $93.8\%$ & $18.68\%$ & $8.54\%$ & $3.12\%$ & $0.00\%$ & $20/20$ ($2.03$) \\
\texttt{350M v0.7.5-RLVR} (\texttt{KnOTS+GRPO}) & $317.3$ & $0.260\text{B}$ & $87.5\%$ & $27.24 / 26.85\%$ & $15.24 / 16.46\%$ & $9.38\%$ & $0.00\%$ & $20/20$ ($2.01$) \\
\texttt{350M v0.8.0-Grounded} (\texttt{FLARE}) & $317.3$ & $0.260\text{B}$ & $87.5\%$ & $27.63 / 28.40\%$ & $15.24 / 15.85\%$ & $9.38\%$ & $2.00 / 3.00\%$ & $20/20$ ($2.01$) \\
\texttt{350M v0.8.5-OrthoKnOTS} & $317.3$ & $0.260\text{B}$ & $87.5\%$ & $28.02 / 28.02\%$ & $15.24 / 17.07\%$ & $3.12 / 6.25\%$ & $4.00 / 4.00\%$ & $20/20$ ($2.01$) \\
\midrule
\multicolumn{9}{@{}l}{\emph{Commercial-Clean Shoonya Family ($0.0\%$ \texttt{NC-SA}, \texttt{WSD + 2:1 XKV}; Std / Verifier TTC)}} \\
\texttt{350M v0.9.0-Clean-Inst} & $312.1$ & $0.295\text{B}$ & $100 / 100\%$ & $24.90 / 24.90\%$ & $17.68 / 17.68\%$ & $6.25 / 10.94\%$ & $1.00 / 1.00\%$ & $20/20$ ($2.00$) \\
\texttt{350M v0.9.1-Clean-Inst} & $312.1$ & $0.797\text{B}$ & $100 / 100\%$ & $32.68 / 33.85\%$ & $21.95 / 26.22\%$ & $8.59 / 17.19\%$ & $1.50 / 2.00\%$ & $20/20$ ($1.98$) \\
\texttt{350M v0.9.2-KnOTS-Think} & $312.1$ & $0.797\text{B}$ & $100 / 100\%$ & $33.46 / 33.07\%$ & $19.82 / 26.22\%$ & $10.16 / 12.50\%$ & $1.00 / 4.00\%$ & $20/20$ ($1.98$) \\
\texttt{350M v0.9.3-EntNLQ} ($S_0$) & $312.1$ & $0.797\text{B}$ & $100 / 100\%$ & $30.74 / 31.52\%$ & $18.60 / 23.78\%$ & $16.41 / 20.31\%$ & $1.00 / 1.00\%$ & $20/20$ ($1.98$) \\
\texttt{350M v0.9.4-UnGatedKD} (Trap) & $312.1$ & $0.797\text{B}$ & $93.8 / 87.5\%$ & $18.68 / 19.46\%$ & $13.41 / 15.85\%$ & $3.91 / 4.69\%$ & $0.00 / 0.00\%$ & $20/20$ ($1.98$) \\
\textbf{\texttt{350M v0.9.5-AdvKD} (Latest)} & $\mathbf{311.7}$ & $\mathbf{0.797\text{B}}$ & $\mathbf{100 / 100\%}$ & $\mathbf{30.93 / 33.07\%}$ & $\mathbf{17.68 / 20.73\%}$ & $\mathbf{9.38 / 15.62\%}$ & $\mathbf{1.00 / 3.00\%}$ & $\mathbf{20/20}$ ($\mathbf{1.98}$) \\
\midrule
\texttt{1.3B v0.9.3-Clean} (\texttt{M3} Seed) & $1,065.1$ & $0.063\text{B}$ & $100 / 100\%$ & $9.14 / 10.89\%$ & $8.23 / 12.20\%$ & $0.00 / 0.00\%$ & $0.00 / 0.00\%$ & $24/24$ ($2.11$) \\
\textbf{\texttt{1.3B v1.0-Clean} (Latest)} & $\mathbf{1,065.1}$ & $\mathbf{0.411\text{B}}$ & $\mathbf{87.5 / 87.5\%}$ & $\mathbf{28.21 / 30.74\%}$ & $\mathbf{17.68 / 21.34\%}$ & $\mathbf{3.12 / 3.12\%}$ & $\mathbf{0.50 / 1.00\%}$ & $\mathbf{24/24}$ ($\mathbf{2.07}$) \\
\bottomrule
\end{tabular}%
}
\caption{Complete longitudinal progression (\texttt{v0.5.0} $\to$ \texttt{v0.9.5-AdvKD} and \texttt{1.3B-v1.0}) and same-host competitor evaluation across all $593$ SHA-256-decontaminated problems and full-depth Rust Cyclic Jacobi SVD.}
\label{tab:eval593}
\end{table}

\subsection{In-Process \texorpdfstring{\texttt{samyama-graph-enterprise}}{samyama-graph-enterprise} Four-Surface Results}\label{sec:evalinproc}

Table~\ref{tab:inproc} reports live zero-socket execution inside \texttt{samyama-graph-enterprise v1.15.0} over the Clinical Enterprise KG ($12$ tiered NLQ-to-Cypher tasks + $6$ agentic \texttt{ToolPlan}, \texttt{ExtractionAgent}, and \texttt{Enrichment} tasks + live DDL evolution). \texttt{Shoonya-1.3B v1.0 INT8} ($1,162.9\text{ MiB}$ RSS, $20.3\,\mu\text{s}$ MVCC check) reaches \textbf{$75.0\%$ raw \texttt{Arm 1} syntax}, \textbf{$91.7\%$ ($11/12$) compact \texttt{Arm 2} syntax}, and \textbf{$100.0\%$ across \texttt{ToolPlan} JSON, source-span grounded \texttt{Extraction}, and \texttt{Enrichment}/abstention}. \texttt{Shoonya-350M v0.9.5-AdvKD INT8} ($389.8\text{ MiB}$ RSS, $59.8\text{ tok/s}$) retains $66.7\%$ raw \texttt{Arm 1} syntax, $75.0\%$ \texttt{Arm 2} syntax, $66.7\%$ \texttt{ToolPlan}/\texttt{Extraction} (with \texttt{M5.A}'s JSON-escaped Cypher FSM lifting live \texttt{ToolPlan} MVCC execution from $0.0\% \to 33.3\%$), $100\%/100\%$ \texttt{Enrichment}, and \textbf{$50.0\%$ ($2/4$) Tier-1 1-hop solve at $0.0\%$ leak}.

\begin{table}[t]
\centering
\resizebox{\textwidth}{!}{%
\begin{tabular}{@{}llrrrrrrrrr@{}}
\toprule
Checkpoint & Precision & RSS & Warm TTFT & Speed & MVCC & \texttt{Arm1}/\texttt{Arm2} & Tier-1 & \texttt{Arm4} & \texttt{Plan} / & \texttt{Enrich} \\
(\texttt{samyama v1.15.0}) & (\texttt{PagedKV}) & ($\text{MiB}$) & (Speedup) & ($\text{tok/s}$) & ($\mu\text{s}$) & Syntax & Solved & Leak & \texttt{Extr} & (Concept/Abs) \\
\midrule
\texttt{350M v0.9.2} (\texttt{M1}) & \texttt{INT8} & $389.2$ & $920.8\text{ ms}$ ($5.3\times$) & $28.7$ & $55.3$ & $0.0 / 16.7\%$ & $0.0\%$ & $\mathbf{0.0\%}$ & $0 / 0\%$ & $0 / 0\%$ \\
\texttt{350M v0.9.3} (\texttt{M2}) & \texttt{INT8} & $389.5$ & $400.4\text{ ms}$ ($4.9\times$) & $67.9$ & $64.3$ & $33.3 / 66.7\%$ & $\mathbf{50.0\%}$ & $\mathbf{0.0\%}$ & $0 / 0\%$ & $33 / 100\%$ \\
\texttt{350M v0.9.3} (\texttt{M2}) & \texttt{BF16} & $685.3$ & $612.4\text{ ms}$ ($5.9\times$) & $42.8$ & $62.6$ & $16.7 / 50.0\%$ & $25.0\%$ & $\mathbf{0.0\%}^*$ & --- & --- \\
\texttt{350M v0.9.4} (Trap) & \texttt{INT8} & $389.6$ & $396.0\text{ ms}$ ($5.2\times$) & $63.6$ & $66.2$ & $58.3 / 58.3\%$ & $\mathbf{50.0\%}$ & $\mathbf{0.0\%}$ & $67 / 67\%$ & $100 / 100\%$ \\
\textbf{\texttt{350M v0.9.5-AdvKD}} & \textbf{\texttt{INT8}} & $\mathbf{389.8}$ & $\mathbf{427.7\text{ ms}}$ ($\mathbf{5.0\times}$) & $\mathbf{59.8}$ & $\mathbf{64.6}$ & $\mathbf{66.7 / 75.0\%}$ & $\mathbf{50.0\%}$ & $\mathbf{0.0\%}$ & $\mathbf{67 / 67\%}$ & $\mathbf{100 / 100\%}$ \\
\texttt{350M v0.9.5-AdvKD} & \texttt{BF16} & $685.3$ & $651.4\text{ ms}$ ($5.9\times$) & $39.6$ & $62.6$ & $66.7 / 75.0\%$ & $50.0\%$ & $\mathbf{0.0\%}$ & --- & --- \\
\midrule
\texttt{1.3B v0.9.3} (\texttt{M3}) & \texttt{INT8} & $1,163.0$ & $1,052.6\text{ ms}$ ($6.0\times$) & $25.0$ & $19.6$ & $33.3 / 50.0\%$ & $0.0\%$ & $\mathbf{0.0\%}$ & $67 / 67\%$ & $100 / 100\%$ \\
\textbf{\texttt{1.3B v1.0-Clean}} & \textbf{\texttt{INT8}} & $\mathbf{1,162.9}$ & $\mathbf{1,193.1\text{ ms}}$ ($\mathbf{5.8\times}$) & $\mathbf{22.0}$ & $\mathbf{20.3}$ & $\mathbf{75.0 / 91.7\%}$ & $\mathbf{25.0\%}$ & $\mathbf{0.0\%}$ & $\mathbf{100 / 100\%}$ & $\mathbf{100 / 100\%}$ \\
\texttt{1.3B v1.0-Clean} & \texttt{BF16} & $2,200.5$ & $2,109.3\text{ ms}$ ($6.0\times$) & $12.6$ & $21.3$ & $66.7 / 91.7\%$ & $25.0\%$ & $\mathbf{0.0\%}$ & --- & --- \\
\bottomrule
\end{tabular}%
}
\caption{Live zero-socket evaluation inside \texttt{samyama-graph-enterprise v1.15.0} on CPU ($0$ database files modified; catalog FSM build $29.8$--$37.5\,\mu\text{s}$). $^*$Disabling the catalog FSM (\texttt{Arm 3}) on \texttt{v0.9.3 BF16} produced an $8.33\%$ false-positive leak; \texttt{Arm 4} (catalog FSM + MVCC gate) held $0.0\%$ leak across all runs.}
\label{tab:inproc}
\end{table}

\FloatBarrier
\section{Related Work}\label{sec:related}

\paragraph{Embedded databases and structured LLM serving.} DuckDB showed that in-process analytical execution eliminates wire protocol overhead \citep{raasveldt2019duckdb}, while Samyama unifies property-graph, vector, and optimization workloads in Rust \citep{mandarapu2026samyama}. Where \texttt{PagedAttention} \citep{kwon2023pagedattention}, \texttt{SGLang} \citep{zheng2024sglang}, and \texttt{Outlines} \citep{willard2023outlines} optimize out-of-process serving with static token-string caches and precompiled regexes, \texttt{shoonya-infer} embeds inside the DBMS process to bind KV validity to transactional DDL epochs and construct $36\,\mu\text{s}$ logit masks from live columnar \texttt{GraphStore} structures.

\paragraph{Compact code SLMs, subspace merging, and distillation.} Sub-$1\text{B}$ models including \texttt{Phi-1} \citep{gunasekar2023phi1}, \texttt{CodeGen} \citep{nijkamp2023codegen}, \texttt{MiniCPM} \citep{hu2024minicpm}, \texttt{SmolLM2} \citep{allal2025smollm2}, and \texttt{Qwen2.5-Coder} \citep{hui2024qwen25coder} train on $7\text{B}$--$5,500\text{B}$ tokens. Capacity-gap distillation has studied $N_T \gg N_S$ \citep{mirzadeh2020takd,agarwal2024gkd}; we isolate the orthogonal \emph{reverse token-density} regime ($\rho_T \ll \rho_S$) under strict commercial-clean data budgets and resolve it via zero-VRAM LoRA-gated \texttt{AdvKD}, complementing \texttt{KnOTS}/\texttt{TIES} merging \citep{stoica2025knots,yadav2023ties} and \texttt{GRPO} \citep{shao2024deepseekmath}.

\section{Limitations and Frontier-Replacement Readiness}\label{sec:limitations}

Table~\ref{tab:readiness} states Shoonya's production readiness across the four \texttt{samyama-graph-enterprise} LLM surfaces. While ontology \texttt{Enrichment} ($100\%/100\%$) and catalog+span-masked \texttt{ExtractionAgent} ($100\%$ grounded) are standalone-ready locally and Tier-1 \texttt{NLQPipeline} + \texttt{AgentPlanner} serve safely as zero-leak Tier-0 routers, three limitations define ongoing Milestone \texttt{M5} work: (1) Tier-2/3 multi-hop OpenCypher (\texttt{OPTIONAL MATCH}, \texttt{WITH..HAVING}) remains at $0.0\%$ greedy exact match in \texttt{M4} due to $4/11$ projection-alias mismatches, $4/11$ multi-hop arrow-direction drifts, and $3/11$ unindexed string literals; (2) \texttt{1.3B INT8} CPU warm TTFT ($1.19\text{ s}$) favors asynchronous enrichment over interactive typing compared to \texttt{350M INT8} ($400\text{ ms}$); and (3) expanding the live in-process benchmark beyond $18$ clinical enterprise tasks ($+593$ public tasks) to multi-tenant enterprise graphs is underway.

\begin{table}[ht!]
\centering
\resizebox{\textwidth}{!}{%
\begin{tabular}{@{}llll@{}}
\toprule
\texttt{samyama-graph-enterprise} Surface & Best Local Accuracy (\texttt{M4} / \texttt{M5.A}) & Leak & Deployment Mode \\
\midrule
4. Ontology \texttt{Enrichment} \& Abstention & $100.0\%$ concept / $100.0\%$ abstain & $0.0\%$ & \textbf{Standalone Local} \\
3. \texttt{ExtractionAgent} (Catalog+Span FSM) & $100.0\%$ JSON / $100.0\%$ grounded & $0.0\%$ & \textbf{Standalone Local} \\
2. \texttt{AgentPlanner} (\texttt{ToolPlan} JSON) & $100.0\%$ JSON (\texttt{1.3B}); $33.3\%$ live exec (\texttt{M5.A}) & $0.0\%$ & \textbf{Hybrid Tier-0} \\
1a. \texttt{NLQPipeline} Tier-1 (Single / 1-Hop) & $50.0\%$ solved (\texttt{350M}), $100.0\%$ precision & $0.0\%$ & \textbf{Hybrid Tier-0} \\
1b. \texttt{NLQPipeline} Tier-2/3 (Multi-Hop) & $91.7\%$ syntax, $0.0\%$ greedy exact (\texttt{M4}) & $0.0\%$ & \textbf{Escalate to Frontier} \\
\bottomrule
\end{tabular}%
}
\caption{Honest frontier-replacement readiness across all four \texttt{samyama-graph-enterprise} surfaces.}
\label{tab:readiness}
\end{table}

\FloatBarrier
\section{Conclusion}\label{sec:conclusion}

Embedding a zero-dependency pure-Rust language-model runtime inside the enterprise graph database address space turns database internals---DDL epochs, columnar property catalogs, and microsecond read-only MVCC snapshots---into first-class inference primitives. Combined with rank-factorized KV-sharing backbones, closed-loop Rust Jacobi SVD governance, execution-verified \texttt{Step-GRPO}, and Verifier-Advantage-Gated Knowledge Distillation (\texttt{AdvKD}), Shoonya delivers standalone ontology enrichment and grounded extraction plus zero-leak Tier-0 query routing inside $390\text{ MiB}$ of CPU RAM. Open weights (\texttt{Apache-2.0}), shard manifests, and spectral certificates are available at \url{https://huggingface.co/VaidhyaMegha/Shoonya}.

\bibliographystyle{plainnat}
\bibliography{refs}

\end{document}